\documentclass[conference]{IEEEtran}
\IEEEoverridecommandlockouts
\usepackage{cite}
\usepackage{amsmath,amssymb,amsfonts}
\usepackage{algorithm}
\usepackage{algorithmic}
\usepackage{graphicx}
\usepackage{textcomp}
\usepackage{xcolor}

\usepackage{booktabs}
\usepackage{tabularx}
\usepackage{ragged2e}

\def\BibTeX{{\rm B\kern-.05em{\sc i\kern-.025em b}\kern-.08em
    T\kern-.1667em\lower.7ex\hbox{E}\kern-.125emX}}

\begin{document}

\title{Quantum Takes Flight: Two-Stage Resilient Topology Optimization for UAV Networks}

\author{
\IEEEauthorblockN{
\thanks{This work is supported by the Canada Research Chairs Program CRC-2022-00187.}
Huixiang Zhang$^{*}$,
Mahzabeen Emu$^{\dagger\ddagger}$,
Octavia A. Dobre$^{\dagger\ddagger}$
}
\IEEEauthorblockA{
$^{*}$Department of Computer Science, Lakehead University, Thunder Bay, Canada\\
$^{\dagger}$Quantum Communications and Computing Research Center, Memorial University, St. John’s, NL, Canada\\
$^{\ddagger}$Department of Electrical and Computer Engineering, Memorial University of Newfoundland, St. John’s, NL, Canada\\
Emails: hzhan102@lakeheadu.ca, memu@mun.ca, odobre@mun.ca
}
}

\maketitle

\begin{abstract}

Next-generation Unmanned Aerial Vehicle (UAV) communication networks must maintain reliable connectivity under rapid topology changes, fluctuating link quality, and time-critical data exchange. Existing topology control methods rely on global optimization to produce a single optimal topology or involve high computational complexity, which limits adaptability in dynamic environments. This paper presents a two-stage quantum-assisted framework for efficient and resilient topology control in dynamic UAV networks by exploiting quantum parallelism to generate a set of high-quality and structurally diverse candidate topologies. In the offline stage, we formulate the problem as a Quadratic Unconstrained Binary Optimization (QUBO) model and leverage quantum annealing (QA) to parallelly sample multiple high-quality and structurally distinct topologies, providing a rich solution space for adaptive decision-making. In the online stage, a lightweight classical selection mechanism rapidly identifies the most suitable topology based on real-time link stability and channel conditions, substantially reducing the computation delay. The simulation results show that, compared to a single static optimal topology, the proposed framework improves performance retention by 6.6\% in a 30-second dynamic window. Moreover, relative to the classic method, QA achieves an additional 5.15\% reduction in objective value and a 28.3\% increase in solution diversity. These findings demonstrate the potential of QA to enable fast and robust topology control for next-generation UAV communication networks.

\end{abstract}


\section{Introduction}


Effective coordination of multi-Unmanned Aerial Vehicle (UAV) swarms is critical for missions in emergency response, logistics, and infrastructure inspection~\cite{torkzaban2025}. However, maintaining stable and efficient communication links presents a significant research challenge. The high mobility of UAVs and unpredictable environmental factors such as wind gusts, node drift, or temporary line-of-sight obstruction cause frequent changes in the UAV network topology. In such dynamic scenarios, a single optimized topology is often fragile; even a small variation in node position or channel condition can lead to a sharp degradation in network performance \cite{garg2023}. A more resilient strategy is to pre-calculate a set of high-quality and structurally diverse candidate topologies. This approach allows UAVs to rapidly switch to the most suitable configuration with minimal computational cost, enabling continuous adaptation to real-time conditions. However, generating this diverse topology set requires repeatedly solving the UAV network topology optimization problem, which is computationally intensive. Conventional approaches, including exact algorithms, heuristic methods, and deep reinforcement learning, face fundamental limitations in meeting the stringent real-time and onboard resource constraints of dynamic UAV networks \cite{chapnevis2024, baktayan2024}.

To address these challenges, we propose a two-stage quantum-assisted framework that decouples computationally intensive topology optimization from online decision making. In the offline phase, we formulate the UAV topology control problem as a Quadratic Unconstrained Binary Optimization (QUBO) model, which serves as the native input format for quantum annealing (QA). This formulation allows the optimization problem to be directly mapped onto quantum hardware, enabling parallel exploration of complex solution landscapes \cite{lin2025}. Unlike classical methods such as simulated annealing (SA), which explore candidate configurations sequentially, QA can explore many possible topologies in parallel. Through quantum-parallel exploration, QA can escape from locally optimal but fragile structures, such as centralized topologies \cite{arai2023,emu2023}. Our QUBO model incorporates penalty terms to ensure load balancing, reduce node overload risk, and avoid single-point failures. In dynamic UAV networks, such imbalances and structural fragility can quickly amplify as the topology changes, leading to severe communication degradation \cite{jeong2025}. To mitigate these effects, we employ QA with an iterative similarity penalty mechanism to generate a structurally diverse and complementary set of candidate topologies. This diversity provides the online stage with a richer set of options, increasing the likelihood of selecting a topology that best matches the current network conditions. During deployment, a lightweight classical evaluation mechanism rapidly selects the most suitable topology based on real-time link stability and Signal-to-Interference-plus-Noise Ratio (SINR), enabling fast adaptation while maintaining network quality.

\begin{figure}[t]
    \centering
    \includegraphics[width=\columnwidth]{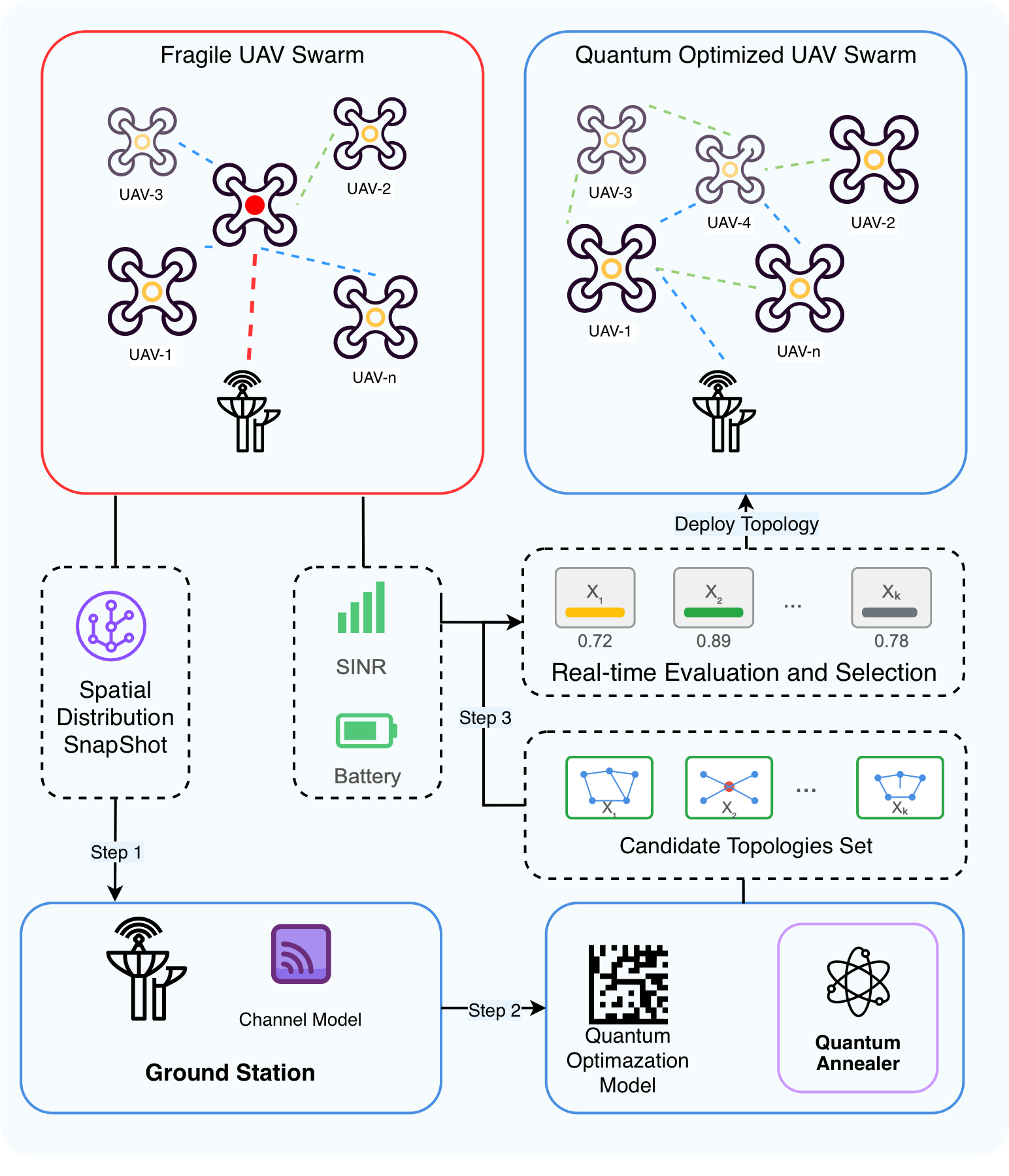}
    \caption{Two-stage topology control framework: offline generation of diverse candidate topologies via QA (Steps 1-2), and online lightweight selection based on real-time network conditions (Step 3).}
    \label{fig:framework}
    \vspace{-0.3cm}
\end{figure}

As shown in Fig. 1, the proposed framework consists of two stages. The left panel depicts a fragile UAV swarm, where a centralized and fragile topology may experience a complete communication breakdown if a single critical node fails. The right panel shows a quantum-assisted swarm with a more balanced and decentralized structure produced by QA. Steps~1 and~2 in Fig.~1 correspond to the offline stage, where the ground station constructs a QUBO model based on the spatial snapshot and channel parameters, then executes QA to produce a set of candidate topologies. Step~3 represents the online stage, in which the UAVs continuously evaluate these candidates using real-time SINR, remaining energy feedback, and select the most suitable topology for deployment, allowing robust and efficient reconfiguration. The major contributions of this paper are summarized as follows:

\textbf{1) Novel Framework:} We propose a two-stage quantum-assisted topology control framework that decouples heavy offline optimization from lightweight online decision-making, enabling scalable and adaptive control in dynamic UAV networks. Further, this framework is aligned with the Software-Defined Networking (SDN) and Open Radio Access Network (O-RAN) architecture, supporting practical deployment.

\textbf{2) Quantum Model:} The offline stage formulates the topology control task as a QUBO problem solved in a QA, which exploits quantum parallelism to explore multiple topologies simultaneously and generate a structurally diverse set of high-quality candidate topologies.

\textbf{3) Lightweight Classical Algorithm:} The online stage employs a simple evaluation mechanism that selects the most suitable topology based on real-time SINR and residual energy, achieving fast adaptation with minimal onboard computation.

\textbf{4) Simulation Results:} Experiments show that our framework improves performance retention by 6.6\% over a static topology baseline, while QA achieves a \textbf{5.15\% reduction in the objective value} and a 28.3\% increase in solution diversity compared to simulated annealing. These findings confirm that integrating offline quantum-based diversity generation with lightweight online evaluation enhances stability and resilience, paving the way for quantum-assisted topology management in next-generation UAV communication networks.

The rest of the paper is organized as follows. Section II reviews related work. Section III presents the system model and the QUBO formulation. Section IV reports on the experimental setup and key results. Section V concludes the paper with limitations and future plans.

\section{Related Work}
Much of the existing literature on UAV network topology control relies on conventional optimization methods. Many studies formulate these problems as Mixed-Integer Linear Programs to find globally optimal solutions for tasks like node placement and trajectory planning \cite{chapnevis2024}. However, these models are NP-hard \cite{chapnevis2024}, and their computational complexity makes them impractical for large and highly dynamic networks. To avoid high computational costs, heuristic and metaheuristic algorithms such as Particle Swarm Optimization and Genetic Algorithms are often applied to tasks like path planning and formation reconfiguration \cite{xiong2025}. While often faster, these methods lack formal convergence guarantees and have unpredictable computation times, making them unreliable for real-time applications with strict deadlines \cite{baktayan2024}. More recently, Deep Reinforcement Learning has been used for UAV topology optimization \cite{xiong2025}, but the substantial computational and power requirements of its deep neural networks conflict with the strict constraints of UAV platforms, reducing mission endurance. The promise of quantum computing, particularly QA, offers a new approach to overcome these classical computational bottlenecks. QA has been successfully applied to several NP-hard problems in networking, such as solving scheduling problems in wireless networks to maximize throughput \cite{emu2023}. However, a central challenge remains: conventional methods are too slow, unreliable, or resource-intensive for dynamic UAV environments \cite{yan2024}, while the application of QA has largely focused on static network problems, and hardware limitations make it unsuitable for direct, real-time control.

Existing approaches either focus on static optimization or rely on heuristic and learning-based methods whose computation time is unpredictable, making them unsuitable for time-critical UAV communication. Hence, a novel framework is needed that can efficiently generate high-quality candidate topologies offline and support rapid adaptation during flight within the operational limits of UAV hardware.

\section{System Model and Proposed Framework}

To bridge the research gap, this section develops the mathematical formulation for our quantum-assisted topology control framework. We first model the UAV communication network as an undirected graph and define the optimization objective that jointly considers throughput maximization and fragility penalization. The main notations in this section are summarized in Table I. The objective is then expressed in the standard QUBO form, enabling direct execution on QA hardware. Finally, we describe how the resulting candidate topologies are integrated into a real-time decision mechanism and discuss the compatibility of this framework with modern SDN/O-RAN architectures.

\begin{table}[htbp]
\centering
\caption{List of Key Notations Used in the Proposed Framework.}
\begin{tabular}{p{0.18\linewidth} p{0.72\linewidth}}
\hline
Symbol & Description \\ \hline
$\mathcal{V}, N$ & Set of UAVs and total number of nodes in the network \\[3pt]
$x_{ij}$ & Binary variable indicating whether the communication link between UAV $i$ and UAV $j$ is activated \\[3pt]
$\mathbf{x}$ & Decision vector representing one feasible UAV topology (set of active links) \\[3pt]
$C_{ij}$ & Channel capacity between UAV $i$ and $j$, calculated using the Shannon capacity formula with an SNR gap approximation for practical modulation schemes \cite{goldsmith1997} \\[3pt]
$T(\mathbf{x})$ & Total achievable throughput of topology (Aggregate Network Capacity) $\mathbf{x}$ \\[3pt]
$F(\mathbf{x})$ & Structural fragility penalty reflecting hub load imbalance \\[3pt]
$\mathcal{N}(k)$ & Neighbor set of node $k$ \\[3pt]

$\alpha, \beta$ & Weight parameters balancing throughput and robustness \\[3pt]
$\mathbf{Q}$ & QUBO matrix constructed from the objective coefficients at the ground control station \\[3pt]
$\mathcal{C}$ & Set of UAV topologies generated by the quantum annealer in the offline stage \\[3pt]
$S_i(t)$ & Real-time utility score of topology $\mathbf{X}_i$ at time $t$ \\[3pt]
$E_l(t)$ & Remaining energy of node $l$ at time $t$ \\[3pt]
$w_{perf}, w_{life}$ & Online weighting parameters balancing throughput and network longevity \\[3pt]
\hline
\end{tabular}
\end{table}

\subsection{Network Model and Objective Function}

We model the UAV communication network as an undirected graph $G=(\mathcal{V},\mathcal{E})$, where $\mathcal{V}$ denotes the set of UAVs with $|\mathcal{V}|=N$, and $\mathcal{E}$ represents all potential communication links between UAVs. Each link $(i,j)\in\mathcal{E}$ is associated with a binary decision variable $x_{ij}\in\{0,1\}$, indicating whether the link is activated. The set of all decision variables forms the decision vector $\mathbf{x}$, which defines one feasible topology.

The quality of a topology is measured by an objective function $C(\mathbf{x})$ that captures the trade-off between the overall throughput and structural robustness:
\begin{equation}
\min_{\mathbf{x}} C(\mathbf{x})=-\alpha\,T(\mathbf{x})+\beta\,F(\mathbf{x}),
\end{equation}
where $T(\mathbf{x})$ represents the proxy for total network throughput and $F(\mathbf{x})$ penalizes topological fragility. The parameters $\alpha$ and $\beta$ control the trade-off between these objectives. These weights are mission dependent and can be tuned according to application priorities. For example, missions requiring continuous data transmission such as UAV monitoring favor larger $\alpha$, while long-endurance or patrol tasks prefer higher $\beta$ to emphasize structural robustness. To enable real-time quantum optimization, we adopt the Aggregate Link Capacity (sum of all active link capacities) as a part of the objective function, defined as:
\begin{equation}
T(\mathbf{x})=\sum_{(i,j)\in\mathcal{E}} C_{ij}\,x_{ij}.
\end{equation}
Here $C_{ij}$ denotes the surrogate link capacity between UAV $i$ and $j$, computed as 
$C_{ij}=B\log_2\!\bigl(1+\gamma_{ij}/\Gamma\bigr)$ with $\Gamma=2$.

The fragility term represents the vulnerability of a topology to node overload and is formulated as follow:
\begin{equation}
F(\mathbf{x})=\sum_{k\in\mathcal{V}}\!\left(\sum_{j\in\mathcal{N}(k)} C_{kj}\,x_{kj}\right)^{\!2}.
\end{equation}
This term penalizes topologies where a small number of UAVs carry disproportionately high traffic loads. The quadratic form enforces balanced traffic distribution, leading to more decentralized and resilient network structures.

\subsection{Stage 1: Offline Strategic Computation}

The objective function can be reformulated into the standard QUBO form $\min_{\mathbf{x}} \mathbf{x}^{T}\mathbf{Q}\mathbf{x}$, where $\mathbf{Q}$ is a symmetric matrix whose elements are derived from the throughput and fragility coefficients. This mapping allows the problem to be directly processed by a quantum annealer.

\begin{algorithm}[!hb]
\caption{Topology Set Generation with Frequency-Based Penalty}
\label{alg:offline}
\begin{algorithmic}[1]
\STATE \textit{Input:} Initial QUBO matrix $\mathbf{Q}_0$, number of rounds $R$, samples per round $k$
\STATE \textit{Output:} Topology set $\mathcal{C}$
\STATE Initialize $\mathcal{C}\leftarrow\emptyset$, $\mathbf{Q}\leftarrow\mathbf{Q}_0$
\FOR{$r=1$ to $R$}
    \STATE $\{\mathbf{X}_{r,1},\dots,\mathbf{X}_{r,k}\} \leftarrow \text{QuantumAnnealer}(\mathbf{Q}, k)$
    \STATE $\mathcal{C}\leftarrow \mathcal{C}\cup\{\mathbf{X}_{r,1},\dots,\mathbf{X}_{r,k}\}$
    \STATE Compute frequency matrix $f_{ij} = \frac{1}{k}\sum_{m=1}^{k} x_{ij}^{(r,m)}$
    \STATE $\mathbf{P}\leftarrow \lambda\,\text{diag}(f_{ij})$
    \STATE $\mathbf{Q}\leftarrow \mathbf{Q}+\mathbf{P}$
\ENDFOR
\STATE \textit{return} $\mathcal{C}$
\end{algorithmic}
\end{algorithm}

As summarized in \textbf{Algorithm~\ref{alg:offline}}, the offline procedure iteratively updates the QUBO matrix to promote exploration and prevent redundant solutions. At the beginning of the process (lines~1–3), the ground control station initializes the QUBO matrix $\mathbf{Q}_0$, specifies the number of annealing rounds $R$, and the number of samples $k$ to be drawn from each run. 

For each round $r \in R$ (lines~4–9), the quantum annealer executes a single annealing process on the current QUBO matrix and outputs $k$ candidate topologies $C = \{\mathbf{X}_{r,1}, \dots, \mathbf{X}_{r,k}\}$, which are merged into the candidate set $\mathcal{C}$. To encourage structural diversity, a similarity penalty matrix $\mathbf{P}$ is constructed according to the occurrence frequency of active links across these $k$ topologies. Specifically, for each link $(i,j)$, its activation frequency is computed as 
\begin{equation}
f_{ij} = \frac{1}{k}\sum_{m=1}^{k} x_{ij}^{(r,m)},
\end{equation}
and a penalty proportional to this frequency is applied to the corresponding diagonal term:
\begin{equation}
P_{ij,ij} \leftarrow P_{ij,ij} + \lambda\, f_{ij}.
\end{equation}
The QUBO matrix is then updated as $\mathbf{Q}\leftarrow\mathbf{Q}+\mathbf{P}$ before the next round, guiding subsequent annealing runs toward unexplored configurations. This iterative quantum–penalty process continues until sufficient distinct topologies are generated, forming the candidate set $\mathcal{C}$ used in the online stage.

\subsection{Stage 2: Real-time Evaluation and Selection}

\begin{algorithm}[h!]
\caption{Real-time Utility Evaluation and Topology Switching}
\label{alg:online}
\begin{algorithmic}[1]
\STATE \textit{Input:} Candidate topology set $\mathcal{C}=\{\mathbf{X}_1,\ldots,\mathbf{X}_k\}$, weights $w_{perf}, w_{life}$
\STATE \textit{Output:} Selected topology for the current deployment $\mathbf{X}_{\mathrm{sel}}$
\STATE Initialize $\mathbf{X}_{\mathrm{current}}$
\REPEAT
    \STATE Collect real-time metrics: $\{\text{SINR}_{jk}(t)\}, \{E_l(t)\}$
    \FORALL{$\mathbf{X}_i\in\mathcal{C}$}
        \STATE Compute $S_i(t)$ using Eq.~(6)
    \ENDFOR
    \STATE $i^* \leftarrow \arg\max_i S_i(t)$
    \STATE $\mathbf{X}_{\mathrm{sel}} \leftarrow \mathbf{X}_{i^*}$
    \IF{$\mathbf{X}_{\mathrm{sel}} \neq \mathbf{X}_{\mathrm{current}}$}
        \STATE $\mathbf{X}_{\mathrm{current}} \leftarrow \mathbf{X}_{\mathrm{sel}}$
        \STATE BroadcastSwitchCommand($\mathbf{X}_{\mathrm{current}}$)
    \ENDIF
\UNTIL{termination condition}
\end{algorithmic}
\end{algorithm}

The second stage operates on UAVs during flight and is designed to be lightweight. 
The swarm receives the precomputed candidate set $C = \{\mathbf{X}_{r,1}, \dots, \mathbf{X}_{r,k}\}$ from the offline stage and continuously evaluates these topologies under current network conditions. Because link quality and energy vary rapidly, the online stage focuses on real-time adaptation by incorporating instantaneous SINR and residual energy, whereas the offline stage emphasizes long-term communication stability. As summarized in \textbf{Algorithm~\ref{alg:online}}, the online process executes a periodic evaluation–selection loop that determines the most suitable topology at each decision interval.

At the beginning of each control cycle (lines~1–3), the UAV swarm initializes the current topology $\mathbf{X}_{\mathrm{current}}$ and loads the candidate set $\mathcal{C}$ with two weighting parameters $w_{perf}$ and $w_{life}$, which balance throughput and energy. During each iteration (lines~4–10), the UAVs first collect real-time network metrics, including instantaneous $\text{SINR}_{jk}(t)$ for all links and the remaining energy $E_l(t)$ for each node. 
For every candidate topology $\mathbf{X}_i \in \mathcal{C}$, a utility score is computed as follow:
\begin{equation}
\begin{split}
S_i(t) = & \, w_{perf} \sum_{(j,k)\in\mathbf{X}_i} \log_2(1+\text{SINR}_{jk}(t)) \\
         & + w_{life} \min_{l\in\mathcal{V}(\mathbf{X}_i)} \left( \frac{E_l(t)}{E_{\text{init}}} \right)
\end{split}
\end{equation}
where the first term reflects short-term communication performance and the second term represents normalized network endurance. 
After computing the utility for all candidates, the topology with the highest score is selected:
\begin{equation}
\mathbf{X}_{\mathrm{sel}} = \arg\max_{\mathbf{X}_i \in \mathcal{C}} S_i(t).
\end{equation}

If the selected topology $\mathbf{X}_{\mathrm{sel}}$ differs from the currently deployed one (lines~11–14), a reconfiguration command is broadcast to all UAVs to update their active links accordingly. 
This evaluation and switching strategy allows the swarm to react quickly to variations in interference, mobility, or energy depletion without solving a new optimization problem in flight. The loop continues until a termination condition is satisfied, such as mission completion or network disconnection.

\subsection{Compatible Deployment with SDN/O-RAN}
To ensure deployability, our framework follows the SDN separation of the control plane and user (data) plane. In this model, the ground controller maintains a global view and offloads the computationally intensive QUBO optimization to a quantum annealer. Meanwhile, the UAVs perform only lightweight execution and local measurements. The same workflow maps cleanly to O-RAN. In the \emph{Non-RT RIC} (non-real-time RAN Intelligent Controller, $>1$~sec timescale) hosted in the SMO/cloud, rApps periodically generate and refresh a diverse set of candidate topologies and issue policies or intent via the A1 interface. In the \emph{Near-RT RIC} (near-real-time RIC, $0.01\text{-}1$~sec control loops), edge xApps consume real-time link/SINR feedback over the E2 interface and rapidly select or switch among those candidates. This architecture aligns with the standard O-RAN functional split, which defines the Radio Unit (RU), Distributed Unit (DU), and Centralized Unit (CU). These units are interconnected by the fronthaul, midhaul, and backhaul transport networks. This hierarchical cloud-edge design facilitates the primary goal of our framework: heavy QUBO computation is offloaded to the cloud, while the UAVs perform only low-cost, seconds-scale selection and switching in practice.

\section{Result Analysis}
We evaluate the proposed two-stage framework under a controlled and fully reproducible setup. Unless otherwise stated, all results are averaged over $20$ independent initial deployments and mobility traces per setting. The simulation horizon is 30 sec with a 1 sec time step. Classical optimization is performed using D-Wave's classical SA solver on an Apple M4 Pro CPU, and quantum solutions are obtained from Qboson Quantum-Inspired hardware CPQC-550 configured with a fixed anneal schedule for fair comparison across runs \cite{KaiwuSDK}. Throughout this section, we use Stage 1 (offline) for candidate set generation and Stage~2 (online) for real-time selection. We compare two strategies, adopting a single optimal method as the baseline. This method computes the topology at $t = 0$ using a D-Wave's SA solver that maximizes total throughput without considering network fragility or load imbalance.

\subsection{Solution Quality and Diversity Comparison}

\begin{figure}[htbp]
    \centering
    \includegraphics[width=\columnwidth]{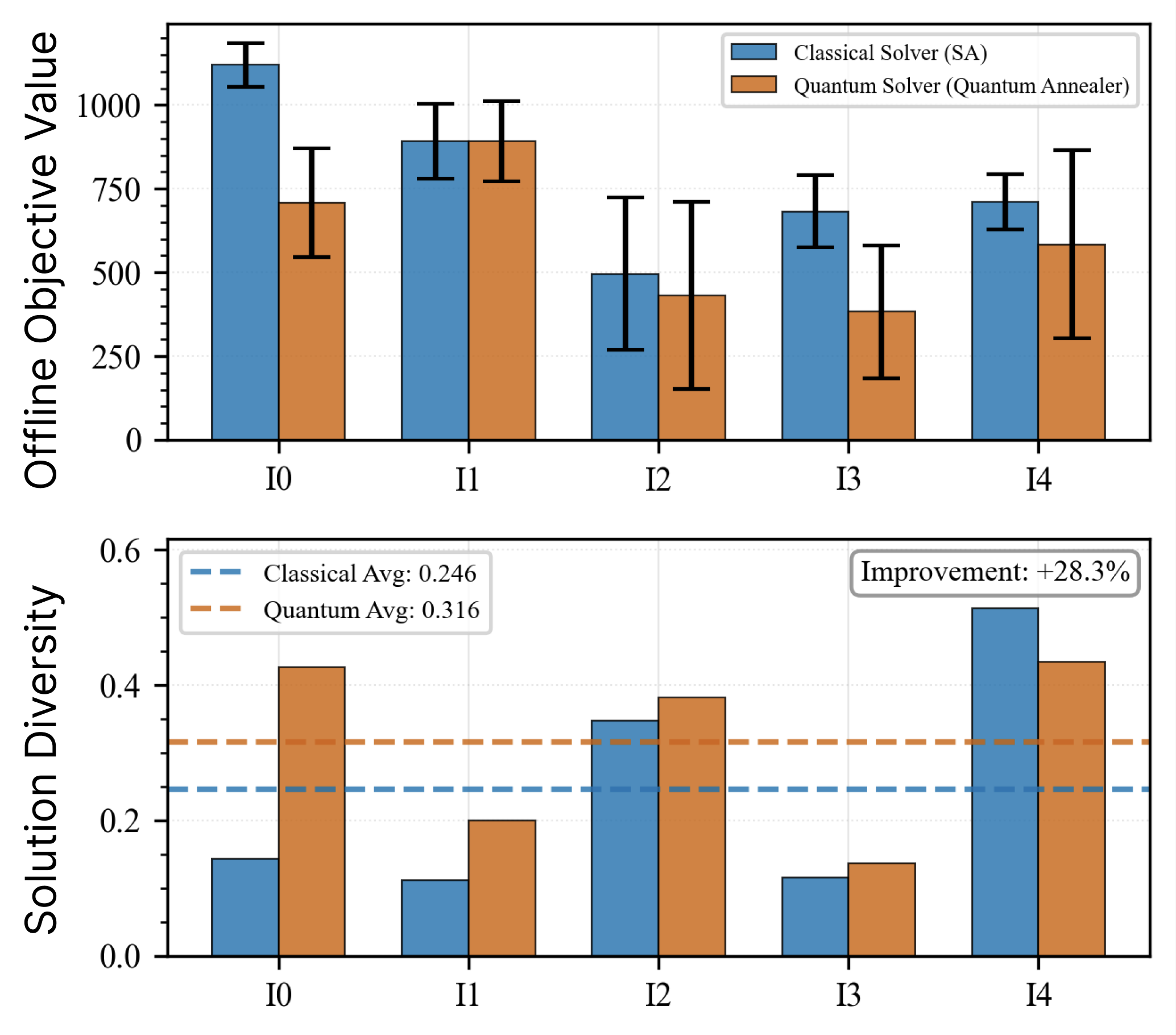}
    \caption{Comparison of offline objective value and solution diversity between QA and SA under five UAV application scenarios, I0 through I4.}
    \vspace{-0.3cm}
\end{figure}
To evaluate the effectiveness of QA in generating high-quality and diverse topologies under the same time budget, we designed five UAV application scenarios that differ in their levels of traffic centralization. Each scenario represents a UAV topology network with distinct communication patterns, ranging from evenly distributed flows to highly centralized control, altering the network’s betweenness centrality distribution from 0.2 to 1.0 \cite{baktayan2024}. Each scenario was repeated three times, and in each run, both QA and SA were allowed to generate ten topologies within an equal computation window. The comparison between QA and SA is presented in Fig.~2. Two metrics are used for evaluation: \textbf{Offline Objective Value}, which refers to the QUBO objective value of each obtained topology, defined as a weighted combination of total throughput and network fragility. A lower value indicates a topology with higher aggregate throughput and more balanced traffic distribution, thus representing better communication performance. \textbf{Solution Diversity} is defined as the average pairwise Hamming distance among all generated topologies, where a higher value reflects richer structural diversity in topology.

As shown in Fig.~2, QA consistently achieves lower offline objective values and higher diversity than SA. Quantitatively, QA improves the average solution diversity by \textbf{28.3\%} and reduces the objective value by \textbf{5.15\%} compared to SA under identical runtime conditions. While SA explores configurations sequentially and often converges to similar local optima, QA leverages quantum superposition and tunneling effects to simultaneously probe multiple low-cost regions of the solution landscape. Consequently, QA produces a larger number of high-quality yet structurally distinct candidate topologies within the same computation window. This diversity advantage yields a richer offline candidate pool for the subsequent online selection stage, enhancing adaptability and overall network resilience in dynamic UAV communication environments.

\begin{figure}[htbp]
    \centering
    \includegraphics[width=\columnwidth]{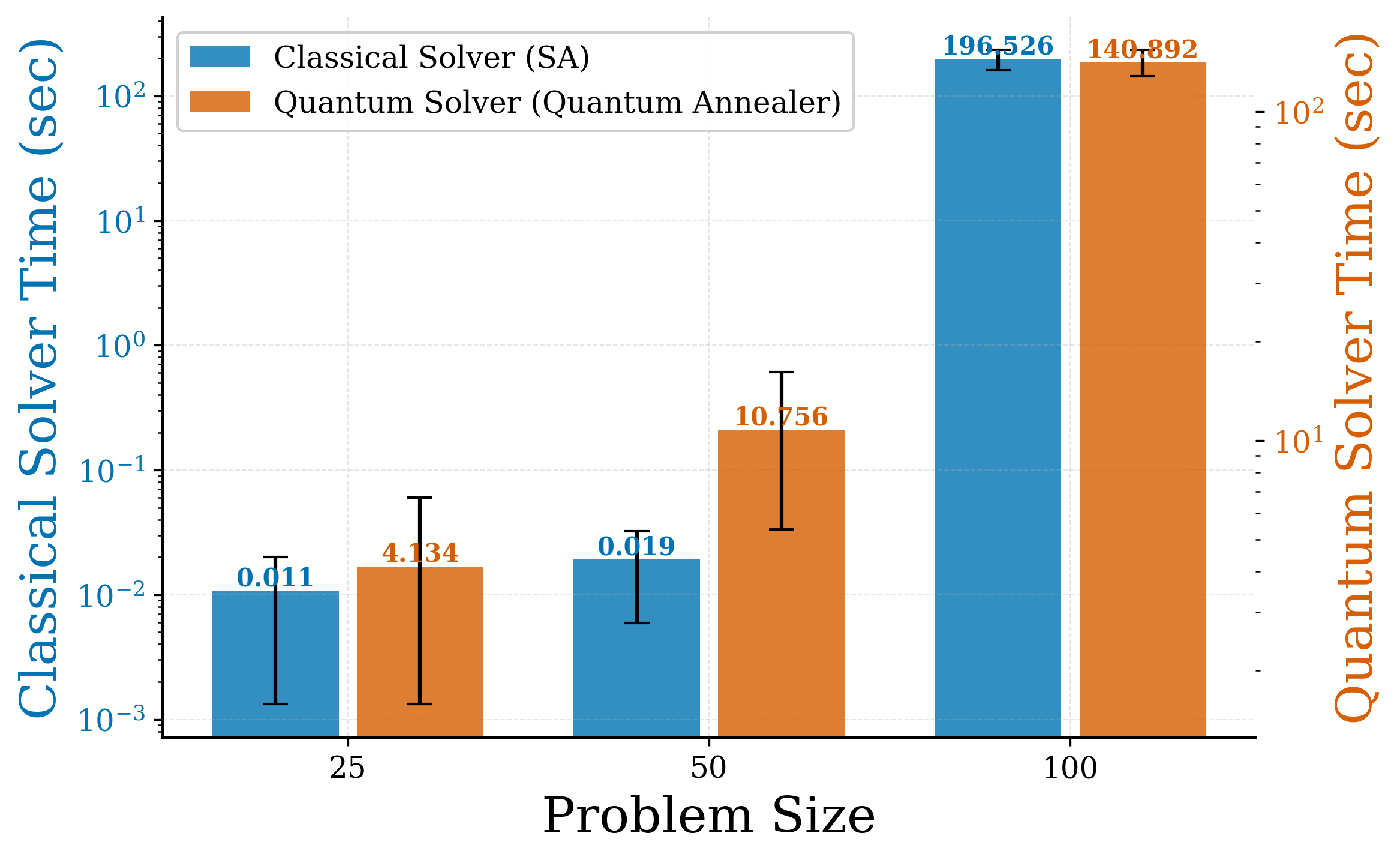}
    \caption{Runtime scaling of a classical solver and a quantum annealer for $N=25, 50, 100$ (log–log). Error bars show standard deviation across runs.}
    \label{fig:solver_comparison}
    \vspace{-0.3cm}
\end{figure}

\subsection{Quantum versus Classical Solver Performance}

To examine the computational scalability for QUBO in Stage 1, we compare SA solver with a quantum annealer across $N \in \{25,50,100\}$ (Fig. 3). For $N=25$, classical SA solver is faster (median 0.011 sec), which is expected for small instances. However, its runtime increases sharply and reaches $196.5$\,s at $N=100$. The quantum annealer shows a more gradual increase, from 4.34 sec at $N=25$ to 140.9 sec at $N=100$, with tight error bars indicating consistent wall-clock behavior across instances. This consistency is important for planning large offline batches: it enables predictable completion of a structurally diverse candidate set. Although classical solvers are faster for small UAV swarms, the quantum annealer exhibits more consistent runtime behavior and produces more diverse solutions as the network size increases. This consistency facilitates large offline batches and better supports the diversity-oriented objective of Stage 1.

\begin{figure}[htbp]
    \centering
    \includegraphics[width=\columnwidth]{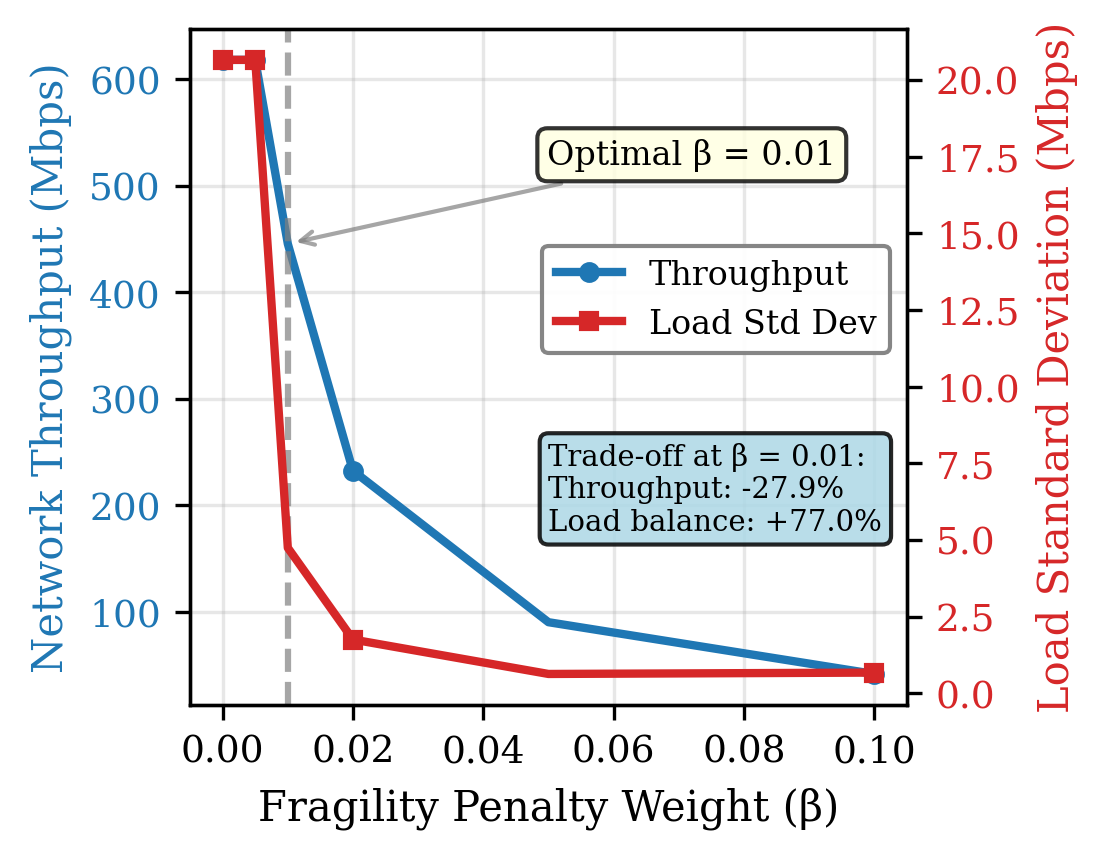}
    \caption{Trade-off between network throughput and load balance as a function of the fragility weight $\beta$. Throughput (blue, left axis) decreases as $\beta$ increases, while load balance (red, right axis), measured by the standard deviation of nodal load, improves accordingly.}
    \vspace{-0.3cm}
\end{figure}

\subsection{Performance-Fragility Trade-off Analysis}
Fig.4 quantifies how $\beta$ balances performance and robustness. As $\beta$ increases, the hub penalty strengthens and both throughput and load imbalance decrease. A clear down appears near $\beta = 0.01$. At this point, the standard deviation of the load improves by $77.0\%$, while the initial throughput experiences a reduction by $27.9\%$. Beyond $\beta = 0.01$, additional robustness gains are small, but the throughput drops more steeply. Therefore, we select $\beta = 0.01$ for the generation of Stage 1 candidates, as it provides an efficient compromise between structural integrity and initial performance.

\begin{figure}[htbp]
    \centering
    \includegraphics[width=\columnwidth]{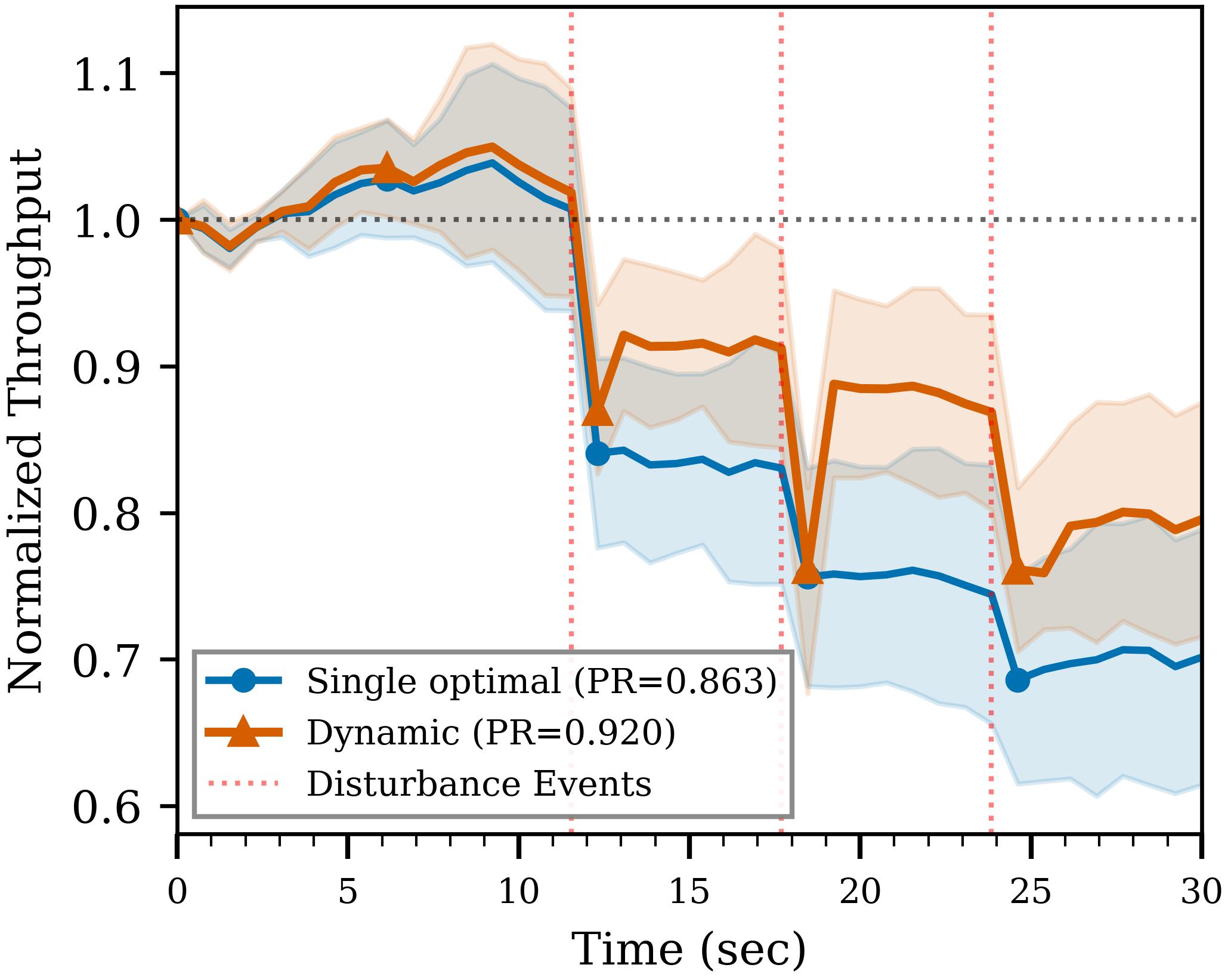}
    \caption{Normalized throughput over 30 s under mobility with mission disturbances. The dashed line at $y=1$ marks the normalization reference, and shaded bands show 95\% confidence intervals over 20 runs. The dynamic two-stage framework ($PR = 0.920$) outperforms the single optimization model ($PR = 0.863$) by 6.6\%, showing faster recovery and greater stability.}
    \label{fig:dynamic_performance}
    \vspace{-0.3cm}
\end{figure}
\subsection{Dynamic Performance Evaluation}
We compare the Single Optimization Model with the proposed Dynamic Two-Stage Framework over a 30 sec mobility window with a 1 sec step. A standard air-to-air channel with standard shadowing is assumed, and link capacities are mapped to network throughput under time-division multiple access scheduling subject to external environmental interference. To reflect realistic UAV operations, 3 disturbance events are injected at approximately 12 sec, 19 sec, and 25 sec, modeled as brief link outages caused by maneuvering or blockage. The single optimization model computes one fixed topology at $t=0$ sec by maximizing total throughput. The dynamic two-stage framework prepares a portfolio of 10 offline candidate topologies and performs lightweight periodic selection with hysteresis and a minimum dwell time, while modeling a short outage for topology switching.
The primary metric is Performance Retention (PR), defined as $\mathrm{PR}=\frac{1}{T}\sum_{t=1}^{T}\frac{\mathrm{thr}(t)}{\mathrm{thr}(0)}$, the time average of throughput ($\mathrm{thr}$) normalized to each method’s own value at $t=0$ sec. As shown in Fig.~\ref{fig:dynamic_performance}, the horizontal (gray) dashed line represents the normalization reference and the shaded bands denote 95\% confidence intervals over 20 runs. The dynamic two-stage framework attains PR $=0.920$, outperforming the Single Optimization Model at PR $=0.863$, which corresponds to a relative improvement of about $6.6\%$, likely to have a greater cumulative improvement on large scale settings. It also recovers faster after mission disturbances and exhibits lower variability, confirming the robustness gained from offline structural diversity combined with online selection.

\section{Conclusion}
This paper presented a two-stage framework to enhance the temporal robustness of communication topologies in dynamic UAV networks. By introducing a fragility penalty into a quantum-compliant QUBO optimization model, we generate inherently resilient decentralized network structures. The offline generation of a diverse topology set, combined with a lightweight online selection mechanism, provides a practical and effective way to achieve sustained network performance without costly in-flight re-optimization. This hybrid approach effectively balances strategic planning with tactical agility, offering a promising solution for robust control in next-generation UAV swarm systems. In future work, we will address scalability and latency via a hierarchical hybrid scheme where classical graph partitioning decomposes large networks exceeding quantum hardware limits, allowing our QUBO-based model to optimize intra-swarm topology while classical methods refine inter-swarm coordination.

\bibliographystyle{IEEEtran}
\bibliography{icc26}

\end{document}